\documentclass[%
 preprint,
superscriptaddress,
 amsmath,amssymb,
 aps,
]{revtex4-2}

\usepackage{graphicx}
\usepackage{dcolumn}
\usepackage{bm}

\begin{document}

\title{Evolution of terahertz third harmonic response across rare-earth nickelate phase-diagram}

\author{Gulloo Lal Prajapati}
\email{gulloo1992@gmail.com}
\author{Igor Ilyakov}
\author{Alexey Ponomaryov}
\affiliation{Helmholtz-Zentrum Dresden-Rossendorf, 01328 Dresden, Germany}

\author{Atiqa Arshad}
\affiliation{Helmholtz-Zentrum Dresden-Rossendorf, 01328 Dresden, Germany}
\affiliation{Institute of Solid State and Materials Physics, TUD Dresden University of Technology, 01069 Dresden, Germany}
\author{Sanjeev Kumar}
\author{Jayaprakash Sahoo}
\author{Dhanvir Singh Rana}
\affiliation{Department of Physics, Indian Institute of Science Education Research Bhopal, Bhopal, Madhya Pradesh 462066, India}

\author{Abdelrahman Azab}
\author{Friedemann Queisser}
\affiliation{Helmholtz-Zentrum Dresden-Rossendorf, 01328 Dresden, Germany}
\author{Ralf Schützhold}
\affiliation{Helmholtz-Zentrum Dresden-Rossendorf, 01328 Dresden, Germany}
    \affiliation{Institut für Theoretische Physik, Technische Universität Dresden, 01062 Dresden, Germany}
\author{Jan-Christoph Deinert}
\affiliation{Helmholtz-Zentrum Dresden-Rossendorf, 01328 Dresden, Germany}


\begin{abstract}

High harmonic generation (HHG) is a sensitive probe for investigating electronic structures and dynamics of materials and a source for attosecond pulses. In particular, HHG with terahertz (THz) light can enable probing of nonlinear responses in correlated materials arising from low-energy many-body interactions. However, THz HHG studies have so far largely focused on topological materials and superconductors, leaving out other potential material systems which could also become efficient THz HHG sources. Here, we report THz third harmonic generation (THG) in rare-earth nickelates -- a prototype material for exploring the Mott insulator-metal transition and related technological applications. We find that the THG amplitude is highly sensitive to the strengths of electronic and magnetic phases of nickelates. In films with sharp phase-transitions, the local maximum and minimum in the temperature-dependent THG amplitude coincide with insulator-metal and magnetic transition temperatures, respectively. While in films with weaker transitions, these features shift toward lower temperatures or even monotonous THG enhancement is observed down to low temperatures. We developed a generalized theory for THz harmonic generation in negative charge-transfer insulators and outlined strategies to enhance the THz nonlinearities further. Our study broadens the scope of THz HHG studies and related applications to strongly correlated materials.   

\end{abstract}

\maketitle

\section{Introduction}
Nonlinear light-matter interaction -- where the incident light itself changes the material properties -- is a powerful tool for controlling and manipulating material properties on an ultrashort timescale \cite{Bloch_2022, Torre_2021}. When the incident light is nearly monochromatic, the nonlinear response of a material can lead to the emission of integer multiples of the driving frequency – a process known as high harmonic generation (HHG). This process is extensively used in producing ultrashort pulses spanning frequencies from infrared to ultraviolet \cite{Corkum_2007, Krausz_2009} and investigating the electronic structure and ultrafast carrier dynamics of materials \cite{Zakse_2012, Hohenleutner_2015, Luu_2018, Bionta_2021, Heide_2022, Zhong_2022}. While these activities were largely limited to the optical regime, the advent of strong terahertz (THz) sources in recent years has led to an extension of HHG research toward these significantly longer wavelengths. THz HHG can be advantageous over its optical counterpart in several aspects, primarily stemming from the lower photon energy and longer wavelength of the THz driving field. The THz band ($\sim$0.1-10 THz) naturally matches the energy scale of the low-energy many-body interactions and collective excitations that define the macroscopic phases of strongly correlated materials \cite{Basov_2011, Ulbricht_2011, Kumar_2020, Yang_2023}. Thus, driving them with intense THz pulses and observing the resulting harmonics can directly probe the nonlinear interplay between charge, spin, orbital and lattice degrees of freedom. THz HHG can be highly sensitive to these low-energy processes, which often get obscured in optical HHG due to electronic transitions and the large thermal effect of the optical photons. Additionally, THz spectroscopy is a phase-sensitive detection technique \cite{Ulbricht_2011} that  allows for the measurement of both the amplitude and phase of the emitted harmonics, providing a complete picture of the material's non-equilibrium response. 

 However, THz HHG studies have so far largely focused on topological materials and superconductors. Topological materials produce highly efficient THz harmonics due to the existence of massless Dirac fermions (charge carriers with a linear energy-momentum dispersion relation) \cite{Hafez_2018, Cheng_2020, Sergey_2021, Kovalev_2021, Jan_2021, Tielroojj_2022, Igor_2023, Atiqa_2023}. For example, in single-layer graphene, generation of THz harmonics up to the $7^{th}$ order with high conversion efficiency was reported \cite{Hafez_2018}. High harmonics in these materials can be tuned by electrical gating \cite{Sergey_2021} and the efficiency can be further enhanced by fabricating metamaterial structures \cite{Jan_2021, Tielroojj_2022}. These high-power harmonics are highly relevant for THz technology such as in 6G communication technology \cite{Kürner_2012, Kürner_2014}, on-chip THz opto-electronics \cite{Ranjan_2020, Ranjan_2025} and THz-rate signal processing \cite{Gulloo_ACS, Tatiana, Nature_Electronics}. THz HHG spectroscopy has been successfully employed to probe the Higgs mode (amplitude fluctuation) in a variety of superconducting materials \cite{Matsunaga_2014, Matsunaga_2017, Chu_2020, Shimano_2020, Isoyama_2021, Kim_2024, Matsumoto_2025}. Higgs mode is a condensed-matter analog of a Higgs boson in particle physics. It does not carry any charge or spin and so, does not respond to linear electromagnetic fields which makes its experimental detection challenging. Since the energy scale of the Higgs mode is comparable to the superconducting gap (a few meV), the coherent interaction between a superconductor and intense THz field results in the nonlinear oscillations of the Higgs mode and thus, can be directly detected in the THz HHG experiment (also known as Higgs spectroscopy in superconductivity community). The THz HHG spectroscopy has also been useful in identifying and segregating other features coexisting with superconductivity such as charge-density fluctuation, charge-density wave, Fano resonance, etc., \cite{Matsunaga_2017, Feng_2023, Chu_2023, Feng_2025}. These studies suggest that THz HHG spectroscopy is highly suitable for probing nonlinear responses arising from low-energy many-body interactions which could also guide synthesizing states of matter with enhanced THz nonlinearities. 

Here, we explore THz nonlinear responses of another complex material system -- rare-earth nickelates (RNiO$_3$ where R is a rare-earth atom). It works as a prototype system to study the Mott insulator-metal transition and explore related technological applications such as modulators, switches, neuromorphic devices, etc.,\cite{Torrance_1992, Middey_2016, Catalano_2018}. Except for LaNiO$_3$, it exhibits three different temperature-dependent phases: a low-temperature antiferromagnetic (AFM) insulating phase, an intermediate-temperature paramagnetic (PM) insulating phase and a high-temperature correlated metallic phase [see Figure~\ref{fig1}(a)]. The metallic phase has orthorhombic crystal structure whereas the insulating phase has monoclinic crystal structure. The structural, electronic and magnetic transitions in nickelates often occur simultaneously and the strengths of different phases and their transition temperatures can be tuned by various means, for example, pressure, epitaxial strain, film thickness, DC field and intense light pulses \cite{Middey_2016, Catalano_2018, Scherwitzl_2010, Monu, Gulloo_2025}. Despite decades of research, the underlying mechanism of the metal-insulator transition (MIT) and ‘cause and effect' in case of simultaneous phase-transitions in nickelates are still topics of active research. Recent discoveries such as coexistence of ﬁrst- and second-order MIT \cite{Post_2018}, onset of a metallic phase in the AFM phase \cite{Mundy} and unusual hysteresis dynamics \cite{Hysteresis} have further raised the interest in this material system. 

In our recent preliminary study, we successfully demonstrated THz third harmonic generation (THG) in this system \cite{Gulloo_2026}. The THG amplitude shows a distinct temperature dependence corresponding to the different nickelate phases. It increases with decreasing temperature in the correlated metallic and AFM insulating phases, while it decreases in the PM insulating phase [Figure~\ref{fig1}(b)]. The sharp rise in the THG amplitude by more than an order of magnitude in the AFM insulating phase is quite remarkable, as one would intuitively expect a decrease in the THG amplitude due to the increased insulating gap. These observations highlight two important points. First, compared to topological materials and superconductors, nickelates have much larger bandgap ($\sim 0.3-1$ eV) which suggests lower band non-parabolicity and weaker free-carrier nonlinearity. Still, THz field is able to drive nickelates into the nonlinear regime and the THG signal is visible in all the three phases. Also note that no HHG is observed in case of the optical drive of the correlated metallic phase \cite{Bionta_2021}. Second, the sharp rise in the THG amplitude with decreasing temperature in the AFM insulating phase suggests potential for nickelates to become an efficient THz HHG material. Since the nickelate phase-diagram possesses complex physics and is highly sensitive to the external conditions, it demands an in-depth investigation how the THz THG evolve across the nickelate phase-diagram under different external conditions. This would enable us to establish the physics of THz harmonic generation as well as explore the factors required for efficient THz harmonic generation in nickelates.

For this purpose, we fabricated several sets of nickelate films to investigate the effect of nickelate phase manipulation (in terms of strength and transition temperature) induced by varying the rare-earth species, epitaxial strain, film thickness and anisotropic strain on the corresponding THz THG responses. In all the experiments, we used narrowband multicycle THz pulses centered at 0.3 THz to drive the nickelate films and measured the emitted THG signal at 0.9 THz (see Section S1, Supporting Information for experimental details). We find that the THG amplitude is highly sensitive to the strength of different phases of nickelate films. In particular, for films exhibiting weak phase-transitions, the temperature dependence of the THG amplitude significantly deviates from the previous observations \cite{Gulloo_2026}. Even a monotonous increase in the THG amplitude is observed down to low temperatures in films with weaker phase-transitions, irrespective of nickelate phases. Through analytical model calculations, we established a detailed mechanism of THz nonlinearities in nickelates and negative charge-transfer insulators in general and discussed a few strategies to further increase the THz nonlinearities.

\section{Results}
\subsection{Phase-diagram of RNiO$_3$}

\begin{figure}[htbp]
    \centering
    \includegraphics[width=\textwidth]{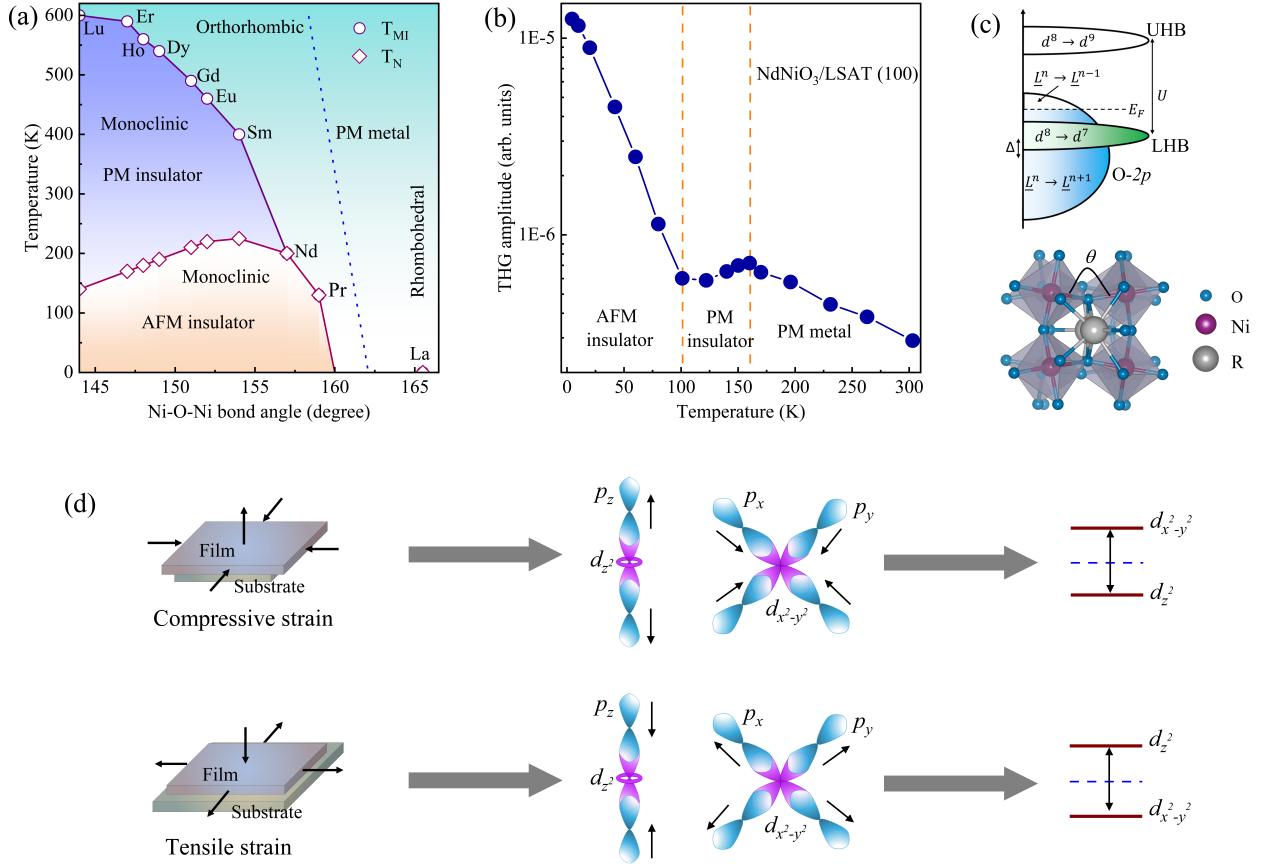}
    \caption{(a) Phase-diagram of bulk RNiO$_3$ as a function of Ni-O-Ni bond angle corresponding to different R atoms. Adapted from \cite{Middey_2016, Catalano_2018,Torrance_1992} (b) Typical temperature dependence of the THz THG amplitude across different nickelate phases as observed in ref. \cite{Gulloo_2026} (c) Top: Single-electron excitation spectra of RNiO$_3$ in terms of charge removal and charge addition within the negative charge-transfer insulator framework. $U$: on-site Coulomb repulsion or energy cost to transfer an electron from occupied $3d$ band (lower Hubbard band, LHB) to unoccupied $3d$ band (upper Hubbard band, UHB). $\Delta$: energy cost to transfer an electron from O-$2p$ band to Ni-$3d$ band. Bottom: Crystal structure of RNiO$_3$. The magnitude of Ni-O-Ni bond angle ($\theta$) depends on the atomic radius of R atoms. The larger the atomic radius, the larger the Ni-O-Ni bond angle. (d) Under epitaxial strain, the film compresses and expands (in the direction of arrows) to acquire in-plane lattice parameters of the underlying substrate. In case of compressive strain, this anisotropic modification reduces the overlapping between O-$p_z$ and Ni-$d_{z^2}$ orbitals while enhances the overlapping between O-$p_{x,y}$ and Ni-$d_{x^2-y^2}$ orbitals. Consequently, the degeneracy of the $e_g$ band is lifted, favoring the electron occupancy in the $d_{z^2}$ orbitals. Opposite occurs in case of tensile strain lowering the energy of the $d_{x^2-y^2}$ orbital.}
    \label{fig1}
\end{figure}

In order to correlate the THz THG response of RNiO$_3$ with its electronic properties, it is essential to first describe the basic properties of RNiO$_3$ and discuss how different external parameters modify these properties. Figure~\ref{fig1}(a) shows the RNiO$_3$ bulk phase-diagram as a function of Ni-O-Ni bond angle corresponding to different RNiO$_3$ members. Within the Zaanen, Sawatzky and Allen (ZSA) scheme \cite{Zaanen_1985, Mizokawa_1994, Bisogni_2016}, RNiO$_3$ is classified as a negative charge-transfer insulator. In these materials, the charge-transfer energy ($\Delta$) corresponding to charge fluctuation  $d_i^n \leftrightarrow d_i^{n+1}\underline{L}$ (where $\underline{L}$ is a ligand hole in the anion valence band) is very small or negative [Figure~\ref{fig1}(c)]. Thus, the ground state of RNiO$_3$ is more accurately represented by $d^8\underline{L}$, i.e., the Ni-$3d$ band is strongly hybridized with the O-$2p$ band and the Ni ions behave more like Ni$^{2+}$ (instead of Ni$^{3+}$). The insulating gap corresponds to the charge fluctuation  $d_i^8\underline{L} + d_j^8\underline{L} \leftrightarrow d_i^8 + d_j^8\underline{L}^2$ and is of $p-p$ character. The effective motion of the holes -- and thus the metallic and insulating nature -- depends critically on the extent of hybridization between the Ni-$3d$ and O-$2p$ orbitals and consequently, on the Ni-O-Ni bond angle. The larger the bond angle, the stronger the hybridization and the higher the hole mobility. The Ni-O-Ni bond angle is proportional to the rare-earth atomic radius. Thus, the overall metallicity decreases and the MIT temperature ($T_\text{MI}$) increases when moving from LaNiO$_3$ toward LuNiO$_3$. On the other hand, the PM to AFM transition temperature ($T_\text{N}$, Néel temperature) begins to saturate after NdNiO$_3$ and gradually decreases because it requires stronger spin-spin correlation strength for the AFM ordering with increasing $T_\text{MI}$.

In thin film form, the epitaxial strain imparted by the underlying substrate is accommodated by modifying the Ni-O bond length and the Ni-O-Ni bond angle \cite{Middey_2016, Catalano_2018}. Thus, the epitaxial strain further modifies the strength and transition temperature of different phases of a given RNiO$_3$. Under compressive strain (when the substrate lattice parameter is smaller than that of the RNiO$_3$), the in-plane Ni-O bond lengths contract and the out-of-plane Ni-O bond lengths elongate leading to the increase of in-plane Ni-O-Ni bond angle and decrease of out-of-plane Ni-O-Ni bond angle [Figure~\ref{fig1}(d)]. This anisotropic modification of the Ni-O bond length and the Ni-O-Ni bond angle increases the overlap between the O-$p_{x,y}$ and the Ni-$d_{x^2-y^2}$ orbitals while reducing the overlap between the O-$p_z$ and the Ni-$d_{z^2}$ orbitals. This raises the energy of the $d_{x^2-y^2}$ orbitals while lowering the energy of the $d_{z^2}$ orbitals. Consequently, the degeneracy of the $e_g$ band is lifted, favoring the electron occupancy in the $d_{z^2}$ orbitals [see the schematic illustration in Figure~\ref{fig1}(d)]. Since the electronic transport largely occurs in-plane, compressive strain increases the overall metallicity of the nickelate film and reduces its $T_\text{MI}$. The opposite occurs under tensile strain (when the substrate lattice parameter is bigger than that of the RNiO$_3$) leading to reduced overall metallicity and increased $T_\text{MI}$. 

\subsection{THG responses of different rare-earth nickelates}

\begin{figure} [htbp]
    \centering
\includegraphics[width=0.7\textwidth]{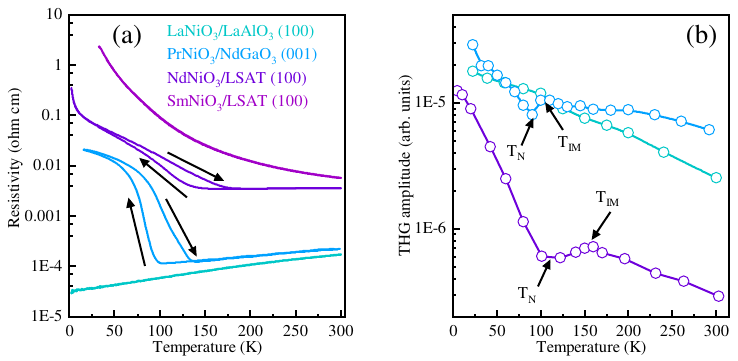}
    \caption{(a) Temperature-dependent resistivity of different RNiO$_3$ films. Arrows show heating and cooling protocols. The corresponding temperature-dependent THG of these films are shown in (b).}
    \label{fig2}
\end{figure}

Figure ~\ref{fig2}(a) shows the temperature-dependent resistivity of LaNiO$_3$, PrNiO$_3$, NdNiO$_3$ and SmNiO$_3$ films. The LaNiO$_3$ film remains metallic and the SmNiO$_3$ film remains insulating in the measured temperature range while PrNiO$_3$ and NdNiO$_3$ films exhibit first-order MIT. Although these films are grown on different substrates which impart different magnitudes of epitaxial strain, and the fact that the epitaxial strain further modifies the overall metallicity and $T_\text{MI}$, their temperature-dependent resistivities align well with the RNiO$_3$ phase-diagram. Figure ~\ref{fig2}(b) shows the temperature-dependent THG amplitude for these films. We could not observe any THG signal from the SmNiO$_3$ film within our experimental resolution. The THG amplitude for other films, as observed in the previous study \cite{Gulloo_2026},  exhibits a distinct temperature-dependence in the three different phases of the nickelates including local maxima and minima at $T_\text{MI}$ and $T_\text{N}$, respectively.

The THG amplitude is the lowest for the NdNiO$_3$ film which is the least conducting among the three films. However, the THG amplitude of the PrNiO$_3$ film in its metallic phase is slightly higher than that of the LaNiO$_3$ film, even though the PrNiO$_3$ film is slightly less conducting. This is due to the slightly higher transmittivity of the PrNiO$_3$ film compared to the LaNiO$_3$ film. This is supported by Figure S1 (Supporting Information) which shows the temperature-dependent THG/FH (fundamental harmonic) ratio, thus removing the effect of different THz transmittivity on the THG amplitude. Further, the THG amplitude of the NdNiO$_3$ film exhibits a stronger temperature dependence than that of the PrNiO$_3$ film. From the temperature-dependent resistivity data of the PrNiO$_3$ film, it is clear that the MIT in this film is relatively more hysteretic and thus, has stronger phase-inhomogeneity (coexistence of metallic and insulating domains) \cite{Hysteresis, Phase-separation}. The phase-inhomogeneity increases the carrier scattering rates \cite{JPCM}, resulting in a less pronounced temperature dependence of the THG amplitude.

\subsection{Role of epitaxial strain} 

\begin{figure}
    \centering
    \includegraphics[width=0.7\textwidth]{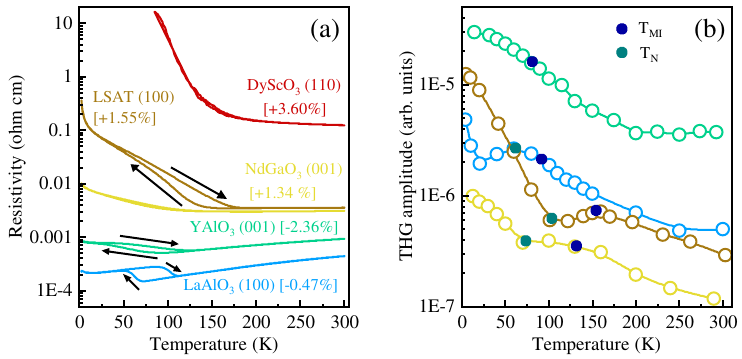}
    \caption{(a) Temperature-dependent resistivity of NdNiO$_3$ films grown on various substrates. The magnitude of epitaxial strain imparted by these substrates are given in the parenthesis. Arrows show resistivity measurements in heating and cooling protocols. The corresponding temperature-dependent THG signals of these films are shown in (b).}
    \label{fig3}
\end{figure}

To explicitly investigate the role of epitaxial strain on the THz THG behavior of nickelates, we fabricated NdNiO$_3$ films on the following substrates: YAlO$_3$ (001), LaAlO$_3$ (100), NdGaO$_3$ (001), LSAT (100) and DyScO$_3$ (110). These substrates impart $-2.61\%$, $-0.45\%$, $+1.37\%$, $+1.58\%$, $+3.38\%$ epitaxial strains, respectively (the negative sign means compressive strain and the positive sign means tensile strain). The temperature-dependent resistivity of these films is shown in Figure~\ref{fig3}(a). The film grown on the DyScO$_3$ (110) substrate remains insulating in the measured temperature range. All the other films exhibit MIT, though the MIT, except for the film grown on LSAT (100) substrate, is quite weak (less than an order of magnitude change in resistivity across the transition) [also see Figure S2]. As expected, the overall resistivity and the $T_\text{MI}$ increase upon going from compressive toward tensile strained films. However, the resistivity of the NdNiO$_3$/YAlO$_3$ (001) film is higher than that of the NdNiO$_3$/LaAlO$_3$ (100) film, even though the magnitude of compressive strain in the former film is larger. One possible reason for this may be, that the increasing compressive strain also increases the density of cation vacancies, leading to increased scattering of the charge carriers \cite{Peng_2016, Aschauer_2013}. The resistivity of oxide films also highly depends on the optimization of the film growth conditions \cite{Liam_2013}.

Figure ~\ref{fig3}(b) shows the temperature-dependent THG amplitude of these films. We did not observe any THG signal from the NdNiO$_3$/DyScO$_3$ (110) film [hence, not plotted in Figure 3(b)]. Clearly, the THG signal is relatively stronger for compressively strained (more metallic) films compared to that for tensile strained (less metallic) films. Interestingly, the local maximum and minimum in the temperature-dependent THG amplitude for NdNiO$_3$/LaAlO$_3$ (100) film do not coincide with its $T_\text{MI}$ and $T_\text{N}$, respectively. Rather, they are significantly shifted toward lower temperatures. For the NdNiO$_3$/YAlO$_3$ (001) film, even a monotonous increase in the THG amplitude is observed down to low temperatures, irrespective of the phase-transition. 

Such temperature dependence of the THG response for different films could be due to the different temperature dependencies of the transmittivities of these films. To exclude the effect of temperature-dependent transmittivity on the observed temperature dependence of the THG response for these films, we plotted the temperature-dependent THG/FH for these films as shown in Figure S3 (Supporting Information). As can be seen, the temperature-dependent THG/FH has a similar shape as the temperature-dependent THG of the respective films. Thus, the temperature dependence of the THz transmission has a negligible influence on the observed temperature-dependence of the THG response.

The observation of a stronger THG signal from compressively strained films is attributed to their higher conductivity. The shift of the local maximum and minimum positions in the THG toward lower temperatures for the NdNiO$_3$/LaAlO$_3$ (100) film can be explained on the basis of the competition between the insulating gap and the orbital-charge correlation below $T_\text{MI}$. While the former will reduce, the latter will increase the THG amplitude \cite{Gulloo_2026, Murakami_2022}. Since the MIT in this film is weak and hysteretic, the initial effect of the orbital-charge correlation on the THG amplitude dominates until the insulating gap becomes large enough to overcome its effect at lower temperatures. Below $T_\text{N}$, spin-charge correlation also comes into play, which has tendency to further increase the THG amplitude \cite{Gulloo_2026, Murakami_2022}. Below 20 K, the net effect of the spin-charge and orbital-charge correlations once again dominates over the effect of the insulating gap. Thus, the THG amplitude starts increasing again with decreasing temperature. In the NdNiO$_3$/YAlO$_3$ (001) film, the net effect of the orbital-charge correlation below $T_\text{MI}$ and the spin-charge correlation below $T_\text{N}$ is so dominant over the effect of the insulating gap that the THG amplitude keeps rising upon cooling down to the lowest measured temperature. By comparing the resistivity and corresponding THG data of the nickelate films, we can safely conclude that the strong decrease in the temperature-dependent THG amplitude in the PM insulating phase occurs only when the MIT is significantly sharp (two-to-three order of magnitude change in the resistivity across the MIT). For a less sharp MIT, only a slight decrease or even an increase in the temperature-dependent THG amplitude is observed down to the low temperatures, irrespective of the phase-transitions.

\subsection{Effect of film thickness}

The thickness of the film can also significantly influence the electronic state of RNiO$_3$. For a compressively strained film, it is observed that the decreasing film thickness increases the insulating nature of the film as well as $T_\text{MI}$ \cite{Peng_2016, Mikheev_2015}. The reverse trend is observed for a tensile strained film. However, for ultrathin films, increased insulating nature and $T_\text{MI}$ can be observed in both the tensile and compressive strain cases due to reduced dimensionality \cite{Scherwitzl_2011}. Such a film thickness-dependent change in the electronic state of RNiO$_3$ arises due to modulation in orbital polarization. Decreasing the film thickness increases orbital polarization (i.e., unequal occupancy of electronic orbitals): compressive strain enhances the occupation of the $d_{z^2}$ orbital while tensile strain enhances the occupation of the $d_{x^2-y^2}$ orbital \cite{Peng_2016, Wu_2013}. Since $T_\text{MI}$ is determined by the extent of overlap between the O-$p_{x,y}$ and Ni-$d_{x^2-y^2}$ orbitals, $T_\text{MI}$ decreases with decreasing thickness of a tensile strained film.  On the other hand, enhanced occupation of the $d_{z^2}$ orbital with decreasing thickness for a compressive strain film means reduced occupation of the $d_{x^2-y^2}$ orbital. Hence, $T_\text{MI}$ increases in this case.

\begin{figure} [htbp]
    \centering
    \includegraphics[width=0.7\textwidth]{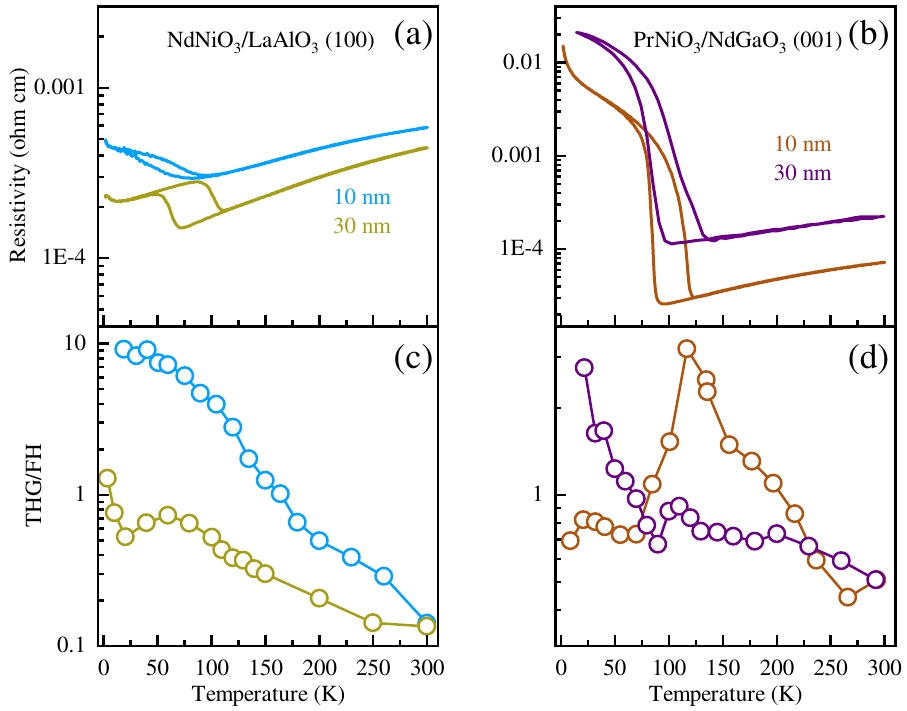}
    \caption{ Temperature-dependent resistivity of (a) NdNiO$_3$/LaAlO$_3$ (compressive strained) films and (b) PrNiO$_3$/NdGaO$_3$ (001) (tensile strained) films of thicknesses 10 nm and 30 nm. The corresponding temperature-dependent THG/FH of these films are shown in (c) and (d), respectively.}
    \label{fig4}
\end{figure}

Figure~\ref{fig4}(a) and (b) show the temperature-dependent resistivity of compressively strained NdNiO$_3$/LaAlO$_3$ (100) and tensile strained PrNiO$_3$/NdGaO$_3$ (001) films of two different thicknesses (10 nm and 30 nm), respectively. The corresponding THG signals from these films are shown in Figure~\ref{fig4}(c) and (d), respectively. Both the NdNiO$_3$/LaAlO$_3$ (100) films exhibit a weak MIT. The THG/FH of the 10 nm film monotonically increases down to low temperatures due to the weaker MIT. On the other hand, both the PrNiO$_3$/NdGaO$_3$ (001) films exhibit a sharp MIT and hence, the THG/FH from them also exhibit the characteristic temperature dependence. Remarkably, the THG/FH trace of the 10 nm film which exhibits the sharpest MIT among all the films described so far, shows the sharpest variation around the $T_\text{MI}$.

\subsection{Effect of anisotropic strain}

\begin{figure} [htbp]
    \centering
    \includegraphics[width=0.8\linewidth]{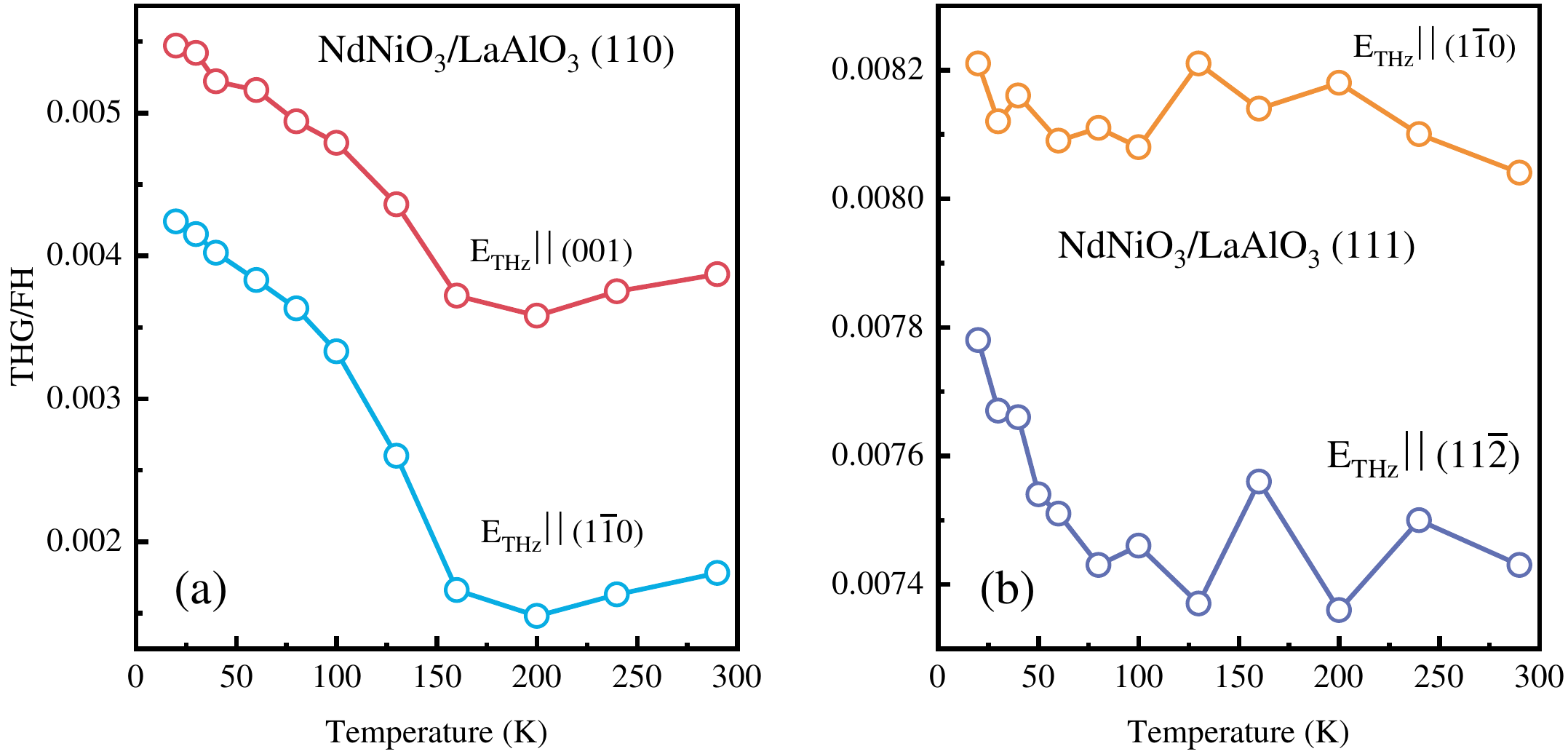}
    \caption{Temperature-dependent THG/FH  measured with THz field polarization along two orthogonal directions of the film plane. (a) NdNiO$_3$/LaAlO$_3$ (110) and (b) NdNiO$_3$/LaAlO$_3$ (111).}
    \label{fig5}
\end{figure}

When the in-plane lattice parameters of the substrate along the two orthogonal axes are unequal, the film experiences different magnitudes of strain along the two orthogonal in-plane directions. In the presence of anisotropic strain, the magnitude of Ni-O hybridization along the two orthogonal directions becomes different. Consequently, the electronic properties of nickelates along the two orthogonal axes also start to differ significantly \cite{Rakesh, Sarmistha}. To investigate the effect of anisotropic strain on the THz THG behavior, we prepared NdNiO$_3$ films on LaAlO$_3$ substrates with (110) and (111) out-of-plane orientations. Thus, for these substrates, the in-plane lattice parameters $a \neq b$. Figure~\ref{fig5}(a) and (b) show the THG signals from the (110) and (111) films, respectively in the two orientations: $E_\text{THz}||(001)$ and $E_\text{THz}||(1\overline{1}0)$ for the (110) film and when $E_\text{THz}||(1\overline{1}0)$ and $E_\text{THz}||(1\overline{1}2)$ for the (111) film. There is a clear difference in the THG/FH magnitudes at any temperature for the two orientations. As expected, the THG/FH magnitude is higher for the orientation with a higher magnitude of compressive strain i.e., for $E_\text{THz}||(001)$ and $E_{THz}||(1\overline{1}0)$ for the (110) and (111) films, respectively.

\subsection{Theoretical model}

To further strengthen our understanding about the THz nonlinear response of RNiO$_3$ or negative charge-transfer insulators in general, we performed analytical model calculations using a tight-binding model Hamiltonian. Details of the calculations are presented in Section S2 (Supporting Information). For the metallic phase, we consider a simple Hamiltonian that describes non-interacting electrons on a regular lattice. The interactions among electrons in the real correlated metallic state can be accounted for in this Hamiltonian by considering non-interacting electrons with their renormalized properties (Fermi liquid theory). We find that the generated current between two neighboring lattice sites (responsible for high harmonic emission) in the presence of a THz driving field not only oscillates at the driving frequency but also contains higher harmonics. In the limiting cases of completely full or empty bands, no high harmonic emission occurs. The same applies to a half-filled but completely spin-polarized band. In these cases, the Pauli principle forbids electrons to hop from one site to the next and thus no current can be generated. Apart from these limiting cases, the metallic state does emit high harmonics under the THz driving field. We also find that the HHG strength reduces at higher temperatures, consistent with our experimental results.

The metal-to-insulator transition in nickelates is accompanied by bond disproportionation, in which alternate Ni-O bond lengths become longer and shorter with effective electronic states $d^8 (S=1)$ and $d^8\underline{L}^2 (S=0)$, respectively \cite{Johnston_2014}. In the AFM insulating state, the electron spins align in a $\uparrow\uparrow\downarrow\downarrow$ pattern with wavevector (1/4, 1/4, 1/4) in the pseudocubic notation. We take account of these factors in the Hamiltonian for the insulating state of nickelates. Under the THz driving field, the dynamics governed by this Hamiltonian result in the emission of high harmonics. Importantly, no high harmonic emission occurs in case of ferromagnetic (FM) spin ordering because the Pauli principle prevents electrons from hopping. The paramagnetic state can be visualized as random spin orientations and would thus fall in between the two extreme cases of AFM and FM spin orderings. This also explains why increasing the temperature of the AFM insulating state lowers the THG amplitude. As the fraction of spin configurations different from AFM spin ordering increases with increasing temperature, the resultant THG amplitude decreases. Another important point is that the figure of merit which determines the relative strength of higher harmonics is the ratio $q\textbf{\textit{E}}\cdot\textbf{\textit{r}}_{\mu\nu}/\omega$ where $q$ is the electronic charge, $\textbf{\textit{E}}$ is the driving field, $\textbf{\textit{r}}_{\mu\nu}$ is the distance between two neighboring sites and $\omega$ is the frequency of the driving field. Since the frequency $\omega$ is in the denominator, it is much easier to reach large values (strong-field regime) of this figure of merit at THz frequencies than at optical frequencies, for example.

\section{Future prospects}

\subsection{THz HHG as a probe for non-linear responses of strongly correlated materials}

We clearly observe the effect of the nickelate’s phase modulation on the THz THG amplitude: the stronger the modulation, the larger the change in the THG amplitude. This one-to-one correspondence between the nickelate’s properties and the THG response may help establishing THz HHG as a tool for probing the nonlinear responses of many-body interactions in strongly correlated materials. To this end, it would be worthwhile to explore both theoretically and experimentally the THz HHG responses of other material systems such as Mott-Hubbard insulators and positive charge-transfer insulators, as well as states of matter such as AFM spin ordering with a metallic state and FM ordering with an insulating state \cite{arxiv}. Finally, the understanding developed here can be extended, in particular, to other $3d-5d$ transition metal oxides which exhibit numerous interesting phases \cite{Kumar_2020}. Considerable progress has been made in recent years in synthesizing thin films and heterostructures of these oxides with the desired properties. The strong sensitivity of the THz harmonic response could be used to study the complex physics of transition metal oxides which itself is an independent research field.

\subsection{Efficient THz harmonic generation in RNiO$_3$}

The observed effects of the ‘R’ species, epitaxial strain and film thickness on the THz THG responses of nickelate films provide insights into how to further increase the THz THG amplitude in nickelates. Clearly, the THG amplitude is larger in highly conducting nickelate films. In the future, nickelate films with higher charge carrier mobility and/or higher charge carrier density can be fabricated to achieve higher THG efficiency. The charge carrier density in nickelate films can also be increased through electron/hole doping such as fabricating Nd$_{0.97}$Ce$_{0.03}$NiO$_3$ film, in which even metallic AFM phase can be stabilized at low temperatures \cite{Mundy}. Another way to generate additional charge carriers is by near/above bandgap optical excitation (photodoping) of the nickelate films \cite{Monu, Torriss_2018}. Such optically generated charge carriers followed by a THz drive should enhance the THG signal \cite{Wang_2017}. We believe that photodoping can significantly enhance the THG efficiency, particularly below $T_\text{MI}$ where the intrinsic free charge carrier density decreases due to the opening of the insulating gap.

\section{Conclusion}

In summary, we investigated the THz THG responses of the nickelate films under several conditions such as by varying the ‘R’ species, epitaxial strain, film thickness and anisotropic strain. We observed a clear one-to-one correspondence between the nickelate properties and their corresponding THG responses. When the nickelate film exhibits a sharp MIT with a large change in resistivity across the transition, the THG response shows characteristic temperature dependence with local maximum and minimum coinciding with $T_\text{MI}$ and $T_\text{N}$, respectively. As the MIT weakens, the local maximum and minimum shift toward lower temperatures. Finally, for films with a much weaker MIT, the THG amplitude increases monotonically down to low temperatures, irrespective of nickelate phases. Using analytical model calculations, we showed how coupling of charge degree of freedom with other fundamental degrees of freedom influences the charge dynamics and the resultant harmonic emission in negative charge-transfer insulators. The strong sensitivity of the THz harmonic response to many-body interactions can be utilized to probe the nonlinear responses in the strongly correlated materials. In the future, it would be desirable to explore the THz harmonic responses of other states of matter in strongly correlated systems. Similarly, several strategies can be employed to further increase the THz THG efficiency in nickelates. We believe that our study will encourage further research in this area and extend THz harmonics research to a broad range of material classes, particularly correlated systems.

\medskip
\textbf{Supporting Information} 

Supporting Information is available from the Wiley Online Library or from the author.

\medskip
\textbf{Acknowledgements}

We thank Dr. Ankit Dulat for useful discussions. This work is funded by the Deutsche Forschungsgemeinschaft (DFG, German Research Foundation) through the Collaborative Research Center SFB 1242 “Nonequilibrium dynamics of condensed matter in the time domain” (Project-ID 278162697). J.-C. D. and G. L. P. acknowledge support from BMBF Verbundprojekt 05K2022 - Tera-EXPOSE. Parts of this research were carried out at ELBE at the Helmholtz-Zentrum Dresden - Rossendorf e. V., a member of the Helmholtz Association.

\end{document}